\documentstyle[11pt]{article}

\newcommand\dati[3]{$#1$&$#2$&$#3$\\ \hline}

\def\capDir{{\bf Scalar field with Dirichlet boundary conditions.}
Values of the zeta function at $s=-1/2$, 
for the inside and the outside regions of a spherical shell, and
values of the Casimir energy. 
The presence of the cutoff $\epsilon$ 
for all even dimensions is to be noted. In such  cases,
the Casimir energy is divergent and needs to be renormalised.}
\def\twoDir{2&
     \dati{+0.0098540 - 0.0039062/\epsilon} 
          {-0.0084955 - 0.0039062/\epsilon} 
          {+0.0006793 - 0.0039062/\epsilon}
	} 
\def\threeDir{3&
     \dati{+0.0088920 + 0.0010105/\epsilon} 
          {-0.0032585 - 0.0010105/\epsilon} 
          {+0.0028168}
	} 
\def\fourDir{4&
     \dati{-0.0017939 + 0.0002670/\epsilon} 
          {+0.0004544 + 0.0002670/\epsilon} 
          {-0.0006698 + 0.0002670/\epsilon}
	} 
\def\fiveDir{5&
     \dati{-0.0009450 - 0.0001343/\epsilon} 
          {+0.0003739 + 0.0001343/\epsilon} 
          {-0.0002856}
	}
\def\sixDir{6&
     \dati{+0.0002699 - 0.0000335/\epsilon} 
          {-0.0000611 - 0.0000335/\epsilon} 
          {+0.0001044 - 0.0000335/\epsilon}
	}
\def\sevenDir{7&
     \dati{+0.0001371 + 0.0000214/\epsilon} 
          {-0.0000555 - 0.0000214/\epsilon} 
          {+0.0000408}
	} 
\def\eightDir{8&
     \dati{-0.0000457 + 5.228\times10^{-6}/\epsilon} 
          {+0.0000101 + 5.228\times10^{-6}/\epsilon} 
          {-0.0000178 + 5.228\times10^{-6}/\epsilon}
	} 
\def\nineDir{9&
     \dati{-0.0000230 - 3.769\times10^{-6}/\epsilon} 
       {+0.0000094 + 3.769\times10^{-6}/\epsilon} 
       {-0.0000068}
}

\def\capNeu{{\bf Scalar field with Neumann boundary conditions (or Robin with the 
choice $\beta=0$).}
Values of the zeta function at $s=-1/2$, 
for the inside and the outside regions of a spherical shell, and corresponding
values of the Casimir energy.} 
\def\twoNeu{2&
    \dati{-0.3446767 - 0.0195312/\epsilon} 
          {-0.0215672 - 0.0195312/\epsilon} 
           {-0.1831220 - 0.0195312/\epsilon}
 }
\def\threeNeu{3&
    \dati{-0.4597174 - 0.0353678/\epsilon} 
          {+0.0120743 + 0.0353678/\epsilon} 
           {-0.2238215}
 }
\def\fourNeu{4&
    \dati{-0.5153790 - 0.0447159/\epsilon} 
          {-0.0060394 - 0.0447159/\epsilon} 
           {-0.2607092 - 0.0447159/\epsilon}
 }
\def\fiveNeu{5&
    \dati{-0.5552071 - 0.0489213/\epsilon} 
          {+0.0030479 + 0.0489213/\epsilon} 
          {-0.2760796}
 }
\def\sixNeu{6&
    \dati{-0.5949395 - 0.0513727/\epsilon} 
          {-0.0128321 - 0.0513727/\epsilon} 
           {-0.3038858 - 0.0513727/\epsilon}
}

\def\capBag{{\bf Massless spinor field
with mixed boundary conditions.}
Values of the zeta function at $s=-1/2$, 
for the inside and the outside regions of a spherical shell, and
values of the Casimir energy. }
\def\twoBag{2&
     \dati{-0.0058312  + 0.0078125/\epsilon}
          {+0.0213677  + 0.0078125/\epsilon}
          {-0.0077683 - 0.0078125/\epsilon} 
               }
\def\threeBag{3&
     \dati{-0.0605944 - 0.0050525/\epsilon}
           {+0.0198217 + 0.0050525/\epsilon}
            {+0.0203863}                       }
\def\fourBag{4&
    \dati{+0.0059074 - 0.0028381/\epsilon}
          {-0.0101965 - 0.0028381/\epsilon}
           {+0.0021445 + 0.0028381/\epsilon}    }
\def\fiveBag{5&
     \dati{+0.0250447 + 0.0025110/\epsilon}
           {-0.0089912 - 0.0025110/\epsilon}
            {-0.0080268}                        }
\def\sixBag{6&
     \dati{-0.0030244 + 0.0011715/\epsilon}
           {+0.0046183 + 0.0011715/\epsilon}
            {-0.0007969 - 0.0011715/\epsilon}    }
\def\sevenBag{7&
     \dati{-0.0108618 - 0.0011745/\epsilon}
           {+0.0040247 + 0.0011745/\epsilon}
            {+0.0034186}                        }

\def\capGS{{\bf Massless spinor field 
with global spectral boundary conditions.}
Values of the zeta function at $s=-1/2$, 
for the inside and the outside regions of a spherical shell, and corresponding
values of the Casimir energy.} 
\def\twoGS{2&
    \dati{-0.0093152 + 0.0319762/\epsilon} 
         {+0.0100172 + 0.0319762/\epsilon} 
         {-0.0003510 - 0.0319762/\epsilon}
 }
\def\threeGS{3&
    \dati{-0.1710212 - 0.0037705/\epsilon} 
         {+0.0019763 + 0.0037705/\epsilon} 
         {+0.0845225}
 }
\def\fourGS{4&
    \dati{+0.0082635 - 0.0118316/\epsilon} 
         {-0.0040473 - 0.0118316/\epsilon} 
         {-0.0021081 + 0.0118316/\epsilon}
 }
\def\fiveGS{5&
    \dati{+0.0680217 + 0.0019471/\epsilon} 
         {-0.0009007 - 0.0019471/\epsilon} 
         {+0.0335605}
 }
\def\sixGS{6&
    \dati{-0.0042224 + 0.0049069/\epsilon} 
         {+0.0017603 + 0.0049069/\epsilon} 
         {+0.0012311 - 0.0049069/\epsilon}
 }
\def\sevenGS{7&
    \dati{-0.0290717 - 0.0009256/\epsilon} 
         {+0.0003983 + 0.0009256/\epsilon} 
         {+0.0143367}
 }
\def\eightGS{8&
    \dati{+0.0020298 - 0.0021417/\epsilon} 
         {-0.0007907 - 0.0021417/\epsilon} 
         {-0.0006196 + 0.0021417/\epsilon}
 }
\def\nineGS{9&
    \dati{+0.0128994 + 0.0004353/\epsilon} 
         {-0.0001787 - 0.0004353/\epsilon} 
         {-0.0063604}
 }

\def\capEF{{\bf Electromagnetic field in a perfectly 
conducting spherical shell.}
Values of the zeta function at $s=-1/2$, 
for the inside and the outside regions of a spherical shell, and corresponding
 values of the Casimir energy. 
Note that in even dimensions, 
in contrast with the scalar field,
the divergences arising from the inside and the
outside energies are different. 
This is due to the fact that 
(only in even dimensions) the $l=0$
mode explicitly contributes to the poles of the
$\zeta$-function, such contribution being absent in the
scalar case.}
\def\twoEF{2&
    \dati{-0.3446767 - 0.0195312/\epsilon}
          {-0.0215672 - 0.0195312/\epsilon}
           {-0.1831220 - 0.0195312/\epsilon}
}
\def\threeEF{3&
    \dati{+0.1678471 + 0.0080841/\epsilon} 
          {-0.0754938 - 0.0080841/\epsilon} 
           {+0.0461767}
}
\def\fourEF{4&
    \dati{+0.5008593 + 0.0231719/\epsilon} 
         {-0.1942082 - 0.0564056/\epsilon} 
           {+0.1533255  -0.0332337/\epsilon}
}
\def\fiveEF{5&
    \dati{+1.0463255 + 0.1838665/\epsilon}                 
          {-0.2981425 - 0.1838665/\epsilon} 
           {+0.3740915}
}
\def\DirBC{\begin{table}\label{Dir0}\begin{center}
\begin{tabular}{|r||l|l||l|}\hline
D&$\zeta(-1/2)$ Inside&$\zeta(-1/2)$ Outside& Casimir Energy\\ 
\hline\hline
\twoDir\threeDir\fourDir\fiveDir\sixDir\sevenDir\eightDir\nineDir
\end{tabular}\caption{\protect\small\capDir}\end{center}\end{table}}
\def\NeuBC{\begin{table}\label{Neu0}\begin{center}
\begin{tabular}{|r||l|l||l|}\hline
D&$\zeta(-1/2)$ Inside&$\zeta(-1/2)$ Outside& Casimir Energy\\ 
\hline\hline
\twoNeu\threeNeu\fourNeu\fiveNeu\sixNeu
\end{tabular}\caption{\protect\small\capNeu}\end{center}\end{table}}
\def\BagBC{\begin{table} \label{Bag12}
\begin{center}\begin{tabular}{|r||l|l||l|}\hline
D&$\zeta(-1/2)$ Inside&$\zeta(-1/2)$ Outside& Casimir Energy\\
\hline\hline
\twoBag\threeBag\fourBag\fiveBag\sixBag\sevenBag
\end{tabular}\caption{\protect\small\capBag}\end{center}\end{table}}
\def\GSBC{\begin{table}\label{GS12}\begin{center}
\begin{tabular}{|r||l|l||l|}\hline
D&$\zeta(-1/2)$ Inside&$\zeta(-1/2)$ Outside& Casimir Energy\\
\hline\hline
\twoGS\threeGS\fourGS\fiveGS\sixGS\sevenGS\eightGS\nineGS
\end{tabular}\caption{\protect\small\capGS}
\end{center}\end{table}}
\def\EF{\begin{table}\label{EF1}\begin{center}
\begin{tabular}{|r||l|l||l|}\hline
D&$\zeta(-1/2)$ Inside&$\zeta(-1/2)$ Outside& Casimir Energy\\
\hline\hline
\twoEF\threeEF\fourEF\fiveEF
\end{tabular}\caption{\protect\small\capEF}
\end{center}\end{table}}

\catcode`\@=11
\@addtoreset{equation}{section}
\def\theequation{\arabic{section}.\arabic{equation}}
\def\appendix{\renewcommand{\thesection}{\Alph{section}}\setcounter{section}{0}
              \renewcommand{\theequation}
            {\mbox{\Alph{section}.\arabic{equation}}}\setcounter{equation}{0}}
\def\R{{\hbox{{\rm I}\kern-.2em\hbox{\rm R}}}}   
\def\H{{\hbox{{\rm I}\kern-.2em\hbox{\rm H}}}}   
\def\N{{\hbox{{\rm I}\kern-.2em\hbox{\rm N}}}}   
\def\C{{\ \hbox{{\rm I}\kern-.6em\hbox{\bf C}}}} 
\def\Z{{\hbox{{\rm Z}\kern-.4em\hbox{\rm Z}}}}   
\renewcommand{\Re}{{\rm Re\,\,}}

\newcommand{\s}[1]{\section{#1}}
\renewcommand{\ss}[1]{\subsection{#1}}
\def\hs{\qquad}               
\def\beq{\begin{eqnarray}}    
\def\eeq{\end{eqnarray}}      
\def\ap{\left.}               
\def\at{\left(}               
\def\aq{\left[}               
\def\ag{\left\{}              
\def\ct{\right)}              
\def\cq{\right]}              
\def\cg{\right\}}             
\def\ii{\infty}

\newcommand{\reals}{\mbox{${\rm I\!R }$}}

\def\beq{\begin{eqnarray}}
\def\eeq{\end{eqnarray}}
\newcommand{\nn}{\nonumber}

\newcommand{\pa}{\partial}

\newcommand{\zb}{\zeta_{{\cal N}}}

\newcommand{\zeb}{\zeta_{{\cal B}}}

\newcommand{\sip}{\frac{\sin (\pi s)}{\pi}}

\newcommand{\al}{\alpha}
\newcommand{\be}{\beta}
\newcommand{\de}{\delta}

\newcommand{\ga}{\gamma}
\newcommand{\la}{\lambda}
\newcommand{\om}{\omega}
\newcommand{\ze}{\zeta}

\newcommand{\Om}{\Omega}
\newcommand{\Si}{\Sigma}

\newcommand{\Ga}{\Gamma}

\begin{document}
\title{CASIMIR ENERGIES FOR SPHERICALLY SYMMETRIC CAVITIES}

\author{{\sc Guido Cognola}\thanks{E-mail address: cognola@science.unitn.it}\\
Dipartimento di Fisica, Universit\`a di Trento\\
and Istituto Nazionale di Fisica Nucleare,\\
Gruppo Collegato di Trento, Italia\\
{\sc Emilio Elizalde}\thanks{E-mail address: elizalde@ieec.fcr.es,
 \ http://www.ieec.fcr.es/cosmo-www/eli.html}\\
Instituto de Ciencias del Espacio (CSIC),\\
Institut d'Estudis Espacials de Catalunya (IEEC/CSIC), \\
Edifici Nexus 201, Gran Capit\`a 2-4, 08034 Barcelona, Spain, \\ and \
Departament ECM i IFAE, Facultat de F\'{\i}sica, \\
Universitat de Barcelona, Diagonal 647,
08028 Barcelona, Spain,  \\
{\sc Klaus Kirsten}\thanks{Present address: Department of Physics and Astronomy,
The University of Manchester, Oxford Road, Manchester, England. E-mail address:
klaus@a13.ph.man.ac.uk}\\
Universit{\"a}t Leipzig, Institut f{\"u}r Theoretische Physik,\\
Augustusplatz 10, 04109 Leipzig, Germany}

\date{\today}
\maketitle

\begin{abstract}
A general calculation of Casimir energies
---in an arbitrary number of dimensions--- for massless quantized fields
in spherically symmetric cavities is carried out. All the most common
situations, including scalar and spinor fields, the electromagnetic field,
and various boundary conditions are treated with the uppermost accuracy.
The final
results are given as analytical, closed expressions in terms
of Barnes zeta functions. A direct numerical evaluation
of the formulas is then performed, which yields highly accurate numbers
 of, in principle, arbitrarily good precision.
\end{abstract}
\newpage

\section{Introduction}
Calculations of Casimir energies in spherically symmetric situations
have attracted the interest of physicists for well over thirty years now.
Since the calculation of Boyer \cite{boye68-174-1764}, who computed
the Casimir energy for a conducting spherical shell and found, to his
surprise, a repulsive force, many different situations in the spherically
symmetric context have been considered. For example, dielectrics were
included
\cite{milt80-127-49} (for the case of plane, parallel surfaces see
\cite{schw78-115-1}) and used later for possible
explanations of sonoluminescence
\cite{milt97-55-4209}--
\cite{bart-u}. Moreover, enormous
interest has been attracted by  the MIT bag model in QCD
\cite{chod74-9-3471}--
\cite{eliz98-31-1743}
and, also, the influence of different boundary conditions has been considered
in detail \cite{bord97-56-4896,lese96-250-448,nest97-57-1284}.

Different methods have been used for dealing with the Casimir effect. Whereas
in the earlier times  the Green function formalism was preferred,
in recent years different approaches ---which make use of contour integral
representations of the involved spectral sums--- are commonplace. Although the
idea for this method, in the specific context of Casimir energies, goes
back to the early days of the subject \cite{davi72-13-1324}, a systematical,
effective and simple application of this approach in various contexts has
only recently been achieved
\cite{bord97-56-4896,lese96-250-448,nest97-57-1284,barv92-219-201,
bord96-37-895,bord96-182-371,bowe99-59-025007}.

The spectral sum which actually appears in the calculation depends on the
regularisation used and may include a cutoff function, to dampen
high frequency contributions \cite{bowe99-59-025007} or,
as in the zeta regularisation technique \cite{eorbz},
complex powers of the eigenvalues
\cite{bord97-56-4896,lese96-250-448,bord96-37-895,bord96-179-215}.
As a result, the details of the computation may differ slightly, e.g.,
in the specific integration contour chosen, but all of them share the
elegance of this method.

In recent contributions we have further developed  the zeta function
technique, in combination with several contour integral representations. Given
the deep connections among zeta functions, heat kernels and functional
determinants \cite{gilk84b}--
\cite{hawk75-43-199},
one advantage of the method
 is that it can be applied, alternatively, to the
calculation of heat kernel coefficients \cite{bord96-37-895},
functional determinants \cite{bord96-182-371,bord96-179-215}
(see also \cite{dowk96-13-1,dowk96-366-89}), as well
as Casimir energies
\cite{eliz98-31-1743}--
\cite{lese96-250-448}.
This clearly shows that zeta functions serve as a unified framework in
different areas of interest.

Here we want to pursue this idea, by using the zeta function framework
in a precise analysis of the
Casimir energy as a function of the dimension of space. Previously
it had been shown, that arbitrary space dimension can be treated elegantly
by making use of Barnes zeta functions, where the dimension can be
considered as a parameter \cite{bord96-182-371,dowk99-}. This has
been applied to the calculation of heat-kernel coefficients
and determinants and it will be here used to study the Casimir energy. Apart
from dealing with arbitrary dimensions, we will introduce scalars,
spinor fields and the electromagnetic field
in a unified way, including the
effects different sets of boundary conditions have on them. In spirit, our
analysis is to be compared with the one of Ambj{\o}rn and Wolfram in Ref.
\cite{ambj83-147-1}, with the difference that the role of
the Epstein zeta function there is here played by the Barnes zeta function.
For a recent analysis on the dimensional dependence of the Casimir energy
for scalar fields with Dirichlet boundary conditions and the electromagnetic
field in the presence of a spherical shell see
\cite{bend94-50-6547,milt97-55-4940}, where
 the space dimension $D$ has been dealt with as a parameter,  and
results for (in principle) all values of real $D$ have been obtained.

The article is organized as follows. In the next section
we briefly recall the definition of Casimir energies in terms of
zeta functions. In Sect. 3, we shortly describe the method
and derive the formulas that are subsequently needed in the
context of Casimir energy calculations \cite{bord96-37-895,bord96-182-371}.
In Sect. 4 we consider the case of a scalar field. For Dirichlet boundary
conditions, the energy in dimensions $D=2$ up to $D=9$ is given there.
The interior and the exterior regions are treated separately.
Afterwards, the changes in the procedure needed for Robin
boundary conditions are explained, and the corresponding
formulas are derived.
Given that the Casimir energy of the electromagnetic field is
determined by using the Casimir energy
of a scalar field satisfying Dirichlet boundary conditions (TE modes)
and a scalar field satisfying Robin boundary conditions (TM modes), these
forms constitute the basis for the electromagnetic  case, and nothing else
needs to be calculated, as will be
 later described in detail (Sect.~6). Before that, Sect. 5 is
devoted to the spinor field. Local bag boundary conditions, as well as
global spectral boundary conditions, are considered.
In the concluding section, Sect. 7, a summary of our main results, as well
as details on how our method is indeed able to yield arbitrary accurate
results, are given.

\s{The Casimir Energy} The Casimir energy of a quantum field
$\Phi(t,\vec x)$ inside a spherical shell is formally given by
\beq E_{Cas}=\frac12\sum_k\om_k,\label{cas2.1} \eeq (we set $\hbar
=c=1$) with the one-particle energies $\om_k=\sqrt{\lambda_k}$
being obtained from \beq -\Delta\phi _k(\vec x )=\lambda_k\phi _k
(\vec x ), \label{cas2.2} \eeq fulfilling also suitable boundary
conditions. The field operator is, in our case,
$A=\partial_t^2-\Delta$, and we have $\Phi(t,\vec x)=e^{-i\om
t}\phi(\vec x)$. The Laplacian $\Delta$ is the one defined inside
or outside the $D=(d+1)$-dimensional ball $B^D=\{\vec
x\in\reals^D|\,\, \| \vec x \| \leq R\}$ and the fields $\phi(\vec
x)$ must satisfy appropriate boundary conditions at $\|\vec
x\|=R$.

The Casimir energy as given by the formal expression (\ref{cas2.1}) is
ill defined and has to be regularised.
In the $\zeta$-function regularisation procedure, one writes
\beq
E_{Cas}=\frac12\mu^{2s}\zeta(s-1/2)|_{s=0}\label{cas2.5}\:,
\eeq
\beq
\zeta(s)=\sum_{\la_k\neq0}\lambda_k^{-s}\:.
\eeq
Here, $\mu$ is an arbitrary parameter with dimensions of mass to yield
the correct dimension for all values of $s$, and $\zeta(s)$
is the $\zeta$-function corresponding to the operator $A$.
In some cases, $E_{Cas}$ will be divergent and, as is known and will be seen
later on, renormalisation ambiguities may remain.

In order to calculate $E_{Cas}$ according to the previous definition,
we need information on the zeta function $\zeta(s)$ in a neighborhood of
 $s=-1/2$.
As we are dealing with operators in flat space, but satisfying boundary
conditions on a $d$-dimensional sphere ($d=D-1$, the boundary of
the $D$-dimensional ball),
the eigenvalues will be implicitly given as the zeros of a polynomial
$\tilde P(\tilde Z_\nu,\tilde Z'_\nu)$
involving Bessel or Hankel functions, according to whether one is considering
the internal or the external domain, respectively.
We will denote the associated zeta functions by $\zeta^{int}(s)$ and
$\zeta^{ext} (s)$.
The total Casimir energy will be the sum of the two
terms, that is
\beq
E_{Cas}=\ap\frac12\mu^{2s}\aq\zeta^{int}(s-1/2)
+\zeta^{ext}(s-1/2)\cq\right|_{s=0}\label{cas25}\:.
\eeq
With just a few modifications,
which involve the phase of the zeta function (see \cite{eejhep} for precise
details), all the considerations above can be extended to
the Dirac operator.
The basic construct turns out to be the zeta function of the square
of the Dirac operator and one encounters a minus sign in Eq. (\ref{cas2.5}).

\s{The method}
\label{method}
The method to be used  here has been developed in the seminal papers
\cite{bord96-37-895,bord97-56-4896,bord96-182-371} and permits
to compute the $\zeta$-function starting from the (indirect) knowledge of
the eigenvalues through an implicit relation of the kind
\beq
\tilde P(\tilde Z_{\nu_l}(\om_{nl}R),\tilde Z'_{\nu_l}(\om_{nl}R))=0\:,
\hs\hs\la_{nl}=\om_{nl}^2\:,
\label{PZ1}
\eeq
where $n,l\geq0$ are the principal and azimuthal quantum numbers respectively.
The degeneracy $d(l)$ of the eigenvalues and the index $\nu_l$ of the Bessel
functions depend on $l$ and on the dimension. Their explicit forms are strictly related
to the fields and the boundary conditions.

The $\zeta$-function can be expressed as an integral
in the complex plane, that is
\beq
\zeta(s)=\sum_{l}\frac{d(l)}{2\pi i}\int_\gamma
k^{-2s}\frac{\partial}{\partial k}
\ln k^{-b\nu_l}\tilde P(\tilde Z_{\nu_l}(k),\tilde Z'_{\nu_l}(k))\:dk\:,
\hs\Re s>\frac{D}2
\:,\label{intJ}\eeq
where the open contour $\gamma$ has to be chosen to run counterclockwise
and to enclose all strictly positive solutions of Eq.~(\ref{PZ1}).
The additional factor $k^{-b\nu_l}$ has been inserted in order to
cancel the pole at the origin,
which is important when deforming the contour in the next step.
In this way $\gamma$ can also include the
origin. Here $b$ is a number which depends on the asymptotic behaviour of
$\tilde P$ at the origin: in our cases it will be $\pm1$ for scalars,
but for spin 1/2
with mixed boundary conditions it turns out to be $\pm2$.

For explicit calculation, it is convenient to write Eq.~(\ref{intJ})
as an integral on the real axis. This can be done by deforming the contour
$\gamma$ to the imaginary axis and by making the substitution $k\to iy$.
In general one has to be careful when deforming the contour in that no
poles in the plane $\Re k \geq 0$ are hit. In fact, for Robin boundary
conditions one has
%
%
\beq
\tilde P(\tilde Z_\nu,\tilde Z'_\nu)=
\alpha\tilde Z_\nu(k)+k\tilde Z'_\nu(k)
=(\alpha-\nu)\tilde Z_\nu(k)+k\tilde Z_{\nu-1}(k)=0\:,
\eeq
which may have solutions for $k\notin\reals$ too, if $\alpha>\nu$.
To avoid these cases, in the following we will consider
$\alpha\leq\nu_0$ only, $\nu_0$ corresponding to the smallest eigenvalue.

With this assumption we can write the $\zeta$-function in the more simple form
\beq
\zeta(s)=\frac{\sin\pi s}\pi\sum_{l=0}^\ii  d(l)
\int_0^\ii y^{-2s}\frac{\partial}{\partial y}
\ln[y^{-b\nu_l}P(\nu_l,y)]\:dy\:,
\label{zetaF}
\eeq
which is valid for $1/2<\Re s<1$
(for details see \cite{bord96-37-895}).
Here, $P(\nu,y)=P(Z_\nu(y),Z'_\nu(y))$ is a polynomial
like $\tilde P$ (aside, possibly, from an irrelevant sign) and the
$Z_\nu(y)=\tilde Z_\nu(iy)$ are
the modified Bessel functions corresponding to $\tilde Z$.
In order to compute the Casimir energy we need the
$\zeta$-function at $s=-1/2$ and so we have to do an analytic
continuation of Eq.~(\ref{zetaF}).

With this aim, let us now employ the asymptotic expansion of the
modified Bessel functions.
For large values of $\nu$, we have \cite{abra70b}
\beq
I_{\nu}(\nu z)\sim\frac1{\sqrt{2\pi\nu}}
\frac{e^{\nu\eta}}{(1+z^2)^{\frac14}}\:\Sigma_1\:,
&&\Sigma_1=\sum_{k=0}^{\infty}\frac{u_k}{\nu^k}\:,
\label{Inu}\\
I'_{\nu}(\nu z)\sim\frac1{\sqrt{2\pi\nu}}
\frac{e^{\nu\eta}(1+z^2)^{\frac14}}{z}\:\Sigma_2\:,
&&\Sigma_2=\sum_{k=0}^{\infty}\frac{v_k}{\nu^k}\:,
\label{I1nu}\\
K_{\nu}(\nu z)\sim\sqrt{\frac\pi{2\nu}}
\frac{e^{-\nu\eta}}{(1+z^2)^{\frac14}}\:\Sigma_3\:,
&&\Sigma_3=\sum_{k=0}^{\infty}(-1)^k\frac{u_k}{\nu^k}\:,
\label{Knu}\\
K'_{\nu}(\nu z)\sim-\sqrt{\frac\pi{2\nu}}
\frac{e^{-\nu\eta}(1+z^2)^{\frac14}}z\:\Sigma_4\:,
&&\Sigma_4=\sum_{k=0}^{\infty}(-1)^k\frac{v_k}{\nu^k}\:,
\label{K1nu}
\eeq
where $\eta=\sqrt{1+z^2}+\ln[z/(1+\sqrt{1+z^2})]$.
The first few coefficients $u_k$ and $v_k$
are listed in \cite{abra70b}, while
higher order coefficients are immediate to obtain by using the recursion relations
\beq
u_{k+1}(t)&=& \frac12t^2(1-t^2)u'_k(t)
+\frac18\int_{0}^{t}(1-5\tau^2)u_k(\tau)\:d\tau\:,\\
v_{k+1}(t)&=&u_{k+1}(t)-\frac12t(1-t^2)u_k(t)-t^2(1-t^2)u'_k(t)\:,\\
t&=&\frac1{\sqrt{1+z^2}}\:,\hs z=\frac{\sqrt{1-t^2}}t\:.
\eeq

As we shall see explicitly in the following, the above behaviour of Bessel
functions permits us to write
\beq
\ln P(\nu,z\nu)\sim
\ln F(\nu,z)+\sum_{n=1}^N\:\frac{D_n(t)}{\nu^n}\:,
\eeq
an expression which is
valid for large values of $\nu$. The function $F$ is related to
the exponential factors in Eqs.~(\ref{Inu})-(\ref{K1nu}), while
the coefficients $D_n(t)$ are related to $\Sigma_k$ and are polynomials in $t$.
More precisely
\beq
D_n(t)=\sum_{k=0}^{2n}\:x_{nk}t^{n+k}\:.
\label{111}
\eeq
Note that when $b=\pm1$, all $x_{nk}$ with odd $k$ vanish.
Of course, $F$, $D_n$ and $x_{nk}$ depend on the specific problem under
 consideration. We will specialize them for every case.

The trick consists now in subtracting the asymptotic behaviour
from the integrand function and in integrating the asymptotic
part, with arbitrary $s$, exactly. We thus get
\beq
\zeta(s)=Z_0(s)+Z(s)+\sum_{n=-1}^{N}A_n(s).
\label{PF}
\eeq
Here,
\beq
Z_0(s)=\de(D-2)\:d[0]\:\sip\int_0^{\infty}
dz\,\,z^{-2s}\frac{\partial}{\partial z}\ln P(0,z)\:
\label{Z0}
\eeq
is the contribution due to $\nu_l=0$, which is present only in
two-dimensions and has to be treated specifically for any case.
$Z(s)$ represents all the other terms with the asymptotic
contributions subtracted,
that is
\beq
Z(s)&=&\sip\sum_{\nu_l>0} d(l)\int_0^{\infty}
dz\,\,(z\nu)^{-2s}\:\:\times\nn\\
&&\hs\hs\frac{\partial}{\partial z}\ag
\ln P(\nu_l,z\nu_l)-\ln F(\nu_l,z)
-\sum_{n=1}^N\frac{D_n(t)}{\nu_l^n}\cg\:,
\label{Zs}\eeq
and $A_n$ are the integrals of the asymptotic part.
They read \cite{bord96-182-371}
\beq
A_n(s)=-\frac1{\Ga(s)}\ze_{\cal N}(s+n/2)
\sum_{k=0}^{2n}x_{nk}\frac{\Ga(s+\frac{n+k}2)}{\Ga(\frac{n+k}2)}\:,
\hs\hs n\geq1\:,
\label{222}
\eeq
\beq
A_{-1}+A_0&=&\sip\sum_{\nu_l>0}d(l)\int_0^{\infty}
dz\,\,(z\nu_l)^{-2s}\frac{\partial}{\partial z}
\ln[(z\nu_l)^{-b\nu_l}F(\nu_l,z)]\nn\\
&=&c_{-1}(s)\ze_{\cal N}(s-1/2)+c_0(s)\ze_{\cal N}(s)\:,
\eeq
\beq
\ze_{\cal N}(s)=\sum_{\nu_l>0}\:d(l)\nu_l^{-2s}\:.
\eeq
Eq.~(\ref{Zs}) is convergent for $(D-2-N)/2<\Re s<1$, thus
 for our aim it is sufficient to subtract
$N=D$ asymptotic terms. This means that with $N=D$ we can directly put $s=-1/2$
in Eq.~(\ref{Zs}) and perform the integral numerically.

As we shall see in the explicit examples,
the base $\zeta$-function, $\zeta_{\cal N}$, can
 be conveniently expressed in
terms of the Barnes zeta function \cite{dowk94-35-4989}, defined
as \cite{barn03-19-374} \beq \zeb(s,a;d)&=&\sum_{\vec
m=0}^{\infty} \frac1{(a+m_1+...+m_d)^s} =\sum_{n=0}^\infty
e_n(d)(a+n)^{-s}\:,\nn\\ e_n(d)&=&\frac{(d+n-1)!}{n!(d-1)!},\nn
\eeq for $\Re s>d$. Obviously, there is an expansion of the kind
\beq e_n(d)=\sum_{\al=0}^{d-1}g_\al(d)(a+n)^\al\nn \eeq and this
yields the expansion of the Barnes zeta function in terms of
Hurwitz zeta functions \cite{barn03-19-374,olve54-247-328} \beq
\zeb(s,a;d)=\sum_{\al=0}^{d-1}g_\al(d)\zeta_H (s-\al,a).
\label{cas2.15} \eeq For example, for $d=2$, we trivially get \beq
\zeb(s,a;2)=\zeta_H(s-1,a)+(1-a)\zeta_H(s,a).\nn \eeq One can show
that the $g_\al(d)$ are connected with the generalised Bernoulli
polynomials \cite{norl22-43-121}. This allows to  determine, in a
direct way, the residues and finite parts of the Barnes zeta
function of the problem at hand. As a result, the asymptotic
contributions in (\ref{PF}) are readily computed.

\section{The scalar field}
\label{S:scalar}

The field equation for this case reads
\beq
-\Delta\phi _k(\vec x )=\lambda_k\phi _k(\vec x ),
\eeq
and has to be supplemented with Dirichlet or Robin  boundary conditions.
Here, $\Delta$ is the Laplace
operator inside or outside the $D=(d+1)$-dimensional ball
and we impose Dirichlet ($\phi(\vec x)|_{|\vec x|=R}=0$)
or Robin ($[\alpha\phi(\vec x)|+\phi'(\vec x)]|_{|\vec x|=R}=0$)
boundary conditions.

In polar coordinates the solutions are
\beq
\phi_{l,m,n} (r,\Omega ) = r^{1-D/2}f_{\nu_1} (\omega_{l,n} r )
Y_{l+D/2}(\Omega )\nn
\eeq
with $\nu_l=l+(D-2)/2$, the
$f_{\nu}(r)$ being Bessel functions and the
$Y_{l+D/2}(\Omega)$ hyperspherical harmonics \cite{erde55b}.

\ss{Scalar field with Dirichlet boundary conditions inside a spherical shell}

In this case, the $f_\nu$ are Bessel functions of the first kind and thus
the eigenvalues $\lambda_{l,n}=\omega^2_{l,n}$ are defined through
\beq
J_{\nu_l}(\omega_{l,n}R)=0\nn
\eeq
and have degeneracies given by
$d(l)=(2l+d-1)\frac{(l+d-2)!}{l! (d-1)!}$.
{}From the last equation, it easily follows that
\cite{dowk94-35-4989,bord96-182-371}
\beq
\zb=\zeb\at2s,\frac{d+1}2;d\ct+\zeb\at2s,\frac{d-1}2;d\ct\:.
\eeq
In this case,
\beq
\tilde P(\tilde Z_\nu(k),\tilde Z'_\nu(k))=J_\nu(k)\:,
\hs\hs P(\nu,z)=I_\nu(z),
\label{f0D}
\eeq
and, as a consequence,
\beq
F(\nu,z)=\frac1{\sqrt{2\pi\nu}}
\frac{e^{\nu\eta}}{(1+z^2)^{\frac14}}\:,
\eeq
\beq
\ln\Sigma_1\sim
\sum_{n=1}^{\infty}\frac{D_n(t)}{\nu^n}.
\eeq
The asymptotic contributions have been calculated to be
\cite{bord96-182-371}
\beq
A_{-1}(s)&=&\frac1{4\sqrt{\pi}}
\frac{\Gamma\left(s-\frac12\right)}
{\Gamma(s+1)}\:\zb\left(s-1/2\right),\nn\\
A_0(s)&=&-\frac14\:\zb(s). \label{cas2.14}
\eeq
The $\zeta$-function for the present situation is obtained by means of
Eqs.~(\ref{111})-(\ref{222}) with the definitions above.

As already anticipated in the previous section, in two dimension we
have an aditional contribution
that has to be  computed explicitly. With this aim, we recall that,
for large $z$,
\beq
I_0(z)=\frac{e^z}{\sqrt{2\pi z}}\left\{1+\frac1{8z}+
  {\cal O} \left( z^{-2}
          \right) \right\} \nn
\eeq and thus we can write \beq Z_0(s)&=&\frac{\sin(\pi
s)}{\pi}\left\{
     \int_0^1dz \,\,z^{-2s}
   \frac{\pa}{\pa z}\ln I_0(z)\right. \nn \\
&&\hs+\int_1^\infty dz\,\,z^{-2s}\left[
    \frac{\pa}{\pa z}\ln I_0(z)-1
       +\frac1{2z}+\frac1{8z^2}\right]\nn\\
 & &\left.\hs\hs
-\frac1{4s}+\frac1{2(s-1/2)}-\frac1{16(s+1/2)}\right\},
\label{Z0sc}
\eeq
where the poles at $s=\pm1/2$ are shown explicitly.
The integrals are now convergent for $s=-1/2$ and can be
computed numerically.

\ss{Scalar field with Dirichlet boundary conditions outside a spherical shell}

Now the radial part of the solutions are Bessel functions of the third kind
(Hankel functions), while $\nu_l$ and $d(l)$ remain the same. Thus, we have
\beq
\nu_l=l+\frac{D-2}2\:,\label{deg00}
\eeq
\beq
d(l)=(2l+d-1)\frac{(l+d-2)!}{l! (d-1)!}\:,
\eeq
\beq
\zb=\zeb\at2s,\frac{d+1}2;d\ct+\zeb\at2s,\frac{d-1}2;d\ct\:,
\label{deg11}
\eeq
\beq
P(\nu,z)=K_\nu(z)\:,\label{p0De1}
\eeq
\beq
F(\nu,z)=\sqrt{\frac\pi{2\nu}}\frac{e^{-\nu\eta}}{(1+z^2)^{\frac14}}\:,
\eeq
\beq
\ln\Sigma_3\sim\sum_{n=1}^{\infty}\frac{D_n(t)}{\nu^n}\:,
\eeq
\beq
A_{-1}(s)&=&-\frac1{4\sqrt{\pi}}\frac{\Gamma\left(s-\frac12\right)}
     {\Gamma(s+1)}\:\zb\left(s-1/2\right),\\
A_0(s)&=&-\frac14\:\zb(s)\:.
\eeq
Owing to the particular relation between $\Si_1$ and $\Si_3$, the coefficients
$D_n(t)$ differ from the corresponding coefficients one has in the internal case
just for the trivial factor $(-1)^n$. The same holds also for the
quantities $A_n(s)$.

In two dimensions we have to consider also the contribution due to $\nu=0$,
which can be obtained with the same arguments as in the previous case,
Eq.~(\ref{Z0sc}). The result is
\beq
Z_0(s) &=&\frac{\sin(\pi s)}{\pi}\left\{
\int_0^1dz \,\,z^{-2s}
   \frac{\pa}{\pa z}\ln K_0 (z)\right. \nn \\
& & \hs +\int_1^\infty dz\,\,z^{-2s}\left[
    \frac{\pa}{\pa z}\ln K_0(z)+1
       +\frac1{2z}-\frac1 {8z^2}\right]\nn\\
 & &\left.\hs\hs
-\frac1{4s}-\frac1{2(s-1/2)}+\frac1{16(s+1/2)}\right\}.
\label{p0De2}
\eeq
The numerical results corresponding to the $\zeta$-functions inside and outside
the shell and the total Casimir energy are reported in Table~1 for the choices
$D=2,...,9$.
\DirBC
For the interior space for $D=2$ and $D=3$ as well as for $D=3$
and the exterior space, our results agree with \cite{lese96-250-448}.
For the whole space in $D=3$ the result is given in
\cite{nest97-57-1284,bend94-50-6547}.

\ss{Scalar field with Robin boundary conditions inside a spherical shell}

In the case of Robin boundary conditions the radial part of the solution is
a combination of Bessel functions with derivatives. For the interior case
we have Bessel functions of the first kind and
their eigenvalues are determined through
\beq
\left(1-\frac D2-\beta\right)J_{\nu_l}(\om_{l,n})
+\om _{l,n}J_{\nu_l} ' (\om _{l,n})=0. \label{cas3.1a}
\eeq
Here we have put $\alpha=1-D/2-\be$ and,
in the spirit of Sect.~\ref{method}, we have to
restrict ourselves to the case $\be\geq1-D/2-\nu_0$.
The choice $\be=0$ represents Neumann boundary conditions.

Also for this case $\nu_l$, $d(l)$ and $\zb$ are given by Eqs.~(\ref{deg00})-(\ref{deg11}),
 now with
\beq
P(\nu,z)=\left(1-\frac D2-\beta\right)
I_\nu(z)+zI'_\nu(z)\:,\nn
\eeq
\beq
F(\nu,z)=\sqrt{\frac\nu{2\pi}}e^{\nu\eta}(1+z^2)^{\frac14}\:,\nn
\eeq
\beq
\ln\at\frac{1-D/2-\beta}{\nu}\:t\Si_1+\Sigma_2\ct
\sim\sum_{n=1}^{\infty}\frac{D_n(t)}{\nu^n}\:,\nn
\eeq
\beq
A_{-1}(s)&=&\frac1{4\sqrt{\pi}}\frac{\Gamma\left(s-\frac12\right)}
     {\Gamma(s+1)}\:\zb\left(s-1/2\right),\nn\\
A_0(s)&=&\frac14\:\zb(s)\nn\:.
\eeq
In two dimensions we have to consider also the contribution
\beq
Z_0(s)&=&\frac{\sin(\pi s)}{\pi}\left\{
     \int_0^1dz \,\,z^{-2s}
\frac{\partial}{\partial z}
\ln(\alpha I_0(z)+z I_0'(z))\right.\nn\\
   &&\hs +\int_1^\infty dz \,\, z^{-2s}
\left[\frac{\partial}{\partial z}
     \ln(\alpha I_0 (z)+z I_0'(z))
-1-\frac1{2z}-\left(\frac38-\alpha\right)\frac1{z^2}\right]
   \nn\\
 & &\left.\hs\hs
+\frac1{4s}+\frac1{2(s-1/2)}+\left(\frac38-\alpha\right)
\:\frac1{2(s+1/2)}\right\}.
\nn\eeq

\ss{Scalar field with Robin boundary conditions outside the spherical shell}

As for Dirichlet, the only difference between the interior and the exterior
case consists in the replacement of Bessel functions with Hankel functions.
Eqs.~(\ref{deg00})-(\ref{deg11}) are valid again, while
\beq
P(\nu,z)=\left(1-\frac D2-\beta\right)
K_\nu(z)+zK'_\nu(z)\:,\nn
\eeq
\beq
F(\nu,z)=\sqrt{\frac{\pi\nu}2}e^{-\nu\eta}(1+z^2)^{\frac14}\:,\nn
\eeq
\beq
\ln\at\frac{1-D/2-\beta}{\nu}\:t\Si_3-\Sigma_4\ct
\sim\sum_{n=1}^{\infty}\frac{D_n(t)}{\nu^n}\:,\nn
\eeq
\beq
A_{-1}(s)&=&-\frac1{4\sqrt{\pi}}\frac{\Gamma\left(s-\frac12\right)}
     {\Gamma(s+1)}\:\zb\left(s-1/2\right),\nn\\
A_0(s)&=&\frac14\:\zb(s)\nn\:.
\eeq
For the $\nu=0$ contribution, we have in this case
\beq
Z_0(s) &=&\frac{\sin(\pi s)}{\pi} \left\{
     \int_0^1dz \,\,z^{-2s}
     \frac{\pa}{\pa z}\ln(\alpha K_0 (z)+zK_0'(z))\right.\nn\\
     &&\hs+ \int_1^\infty dz \,\,z^{-2s}\left[
    \frac{\pa}{\pa z}\ln (\alpha K_0 (z)+z K_0' (z))
+1-\frac1{2z}+\left(\frac38-\alpha\right)\frac1{z^2}\right]
   \nn\\
 & &\left.\hs\hs
+\frac1{4s}-\frac1{2(s-1/2)}-\left(\frac38-\alpha\right)
\:\frac1{2(s+1/2)}\right\}.
\nn\eeq
All numerical results corresponding to Neumann boundary conditions
(or Robin ones with $\be=0$) are exhibited
in Table~2.
\NeuBC
For $D=2$ the result is given in \cite{lese96-250-448}, for $D=3$ in
\cite{nest97-57-1284}.

\s{Spinor field on the $D$-dimensional ball: bag boundary conditions}
\label{S:spinor}

We now consider spinor fields, see \cite{dowk96-13-2911,dowk99-}.
The eigenvalue Dirac equation on the Euclidean $D$-ball is
\beq
-i\Ga^\mu\nabla_\mu\psi_{\pm}=\pm k\psi_{\pm},
\quad\Ga^{(\mu}\Ga^{\nu)}=g^{\mu\nu},
\label{Dirac}
\eeq
and the nonzero modes are separated in polar coordinates,
$ds^2=dr^2+r^2d\Om^2$, in standard fashion to be regular at the origin
($C$ and $A$ are radial normalisation factors),
\beq
\psi_{\pm}^{(+)}&=&{A\over r^{(D-2)/2}}
\left( \begin{array}{c}
{iJ_{n+D/2}(kr)\,Z^{(n)}_+(\Om)} \\
{\pm J_{n+D/2-1}(kr)\,Z^{(n)}_+(\Om)}
                \end{array} \right)  ,\nn\\
\psi_{\pm}^{(-)}&=&{C\over r^{(D-2)/2}}
\left( \begin{array}{c} {\pm J_{n+D/2-1}(kr)\,Z^{(n)}_
-(\Om)}\\
{iJ_{n+D/2}(kr)\,Z^{(n)}_-(\Om)} \end{array} \right)  .
\label{diracmodes}
\eeq
Here the $Z^{(n)}_{\pm}(\Om)$ are well-known spinor modes on the
unit $(D-1)$--sphere
(some modern references are
\cite{awad84-245-161,camp96-15-1,jaro92-213-135})
satisfying the intrinsic equation
\beq
-i\ga^j\widetilde\nabla_j Z^{(n)}_{\pm}=\pm\la_n Z^{(n)}_{\pm},
\label{spheig}
\eeq
where
$$
\la_n=n+{D-1\over2},\quad n=0,1,\ldots\,\,.
$$
For $D\geq 2$,
each eigenvalue is greater than or equal to $1/2$ and has degeneracy
\beq
{1\over2}d_s\,\left( \begin{array}{c}
                  {D+n-2} \\
                    n  \end{array} \right) . \nn
\eeq
The dimension, $d_s$, of $\psi$--spinor space is $2^{D/2}$ if $D$ is
even. For odd $D$ it is $2^{(D+1)/2}$ and has
been doubled in order to implement the boundary conditions.
The projected $\ga$-matrices are given by
\beq
\Ga^r=\left(\matrix{{\bf0}&{\bf 1}\cr{\bf1}&{\bf0}\cr}\right),\quad
\Ga^j=\left(\matrix{{\bf0}&i\ga^j\cr-i\ga^j&{\bf0}\cr}\right),\quad\Ga^5=
\left(\matrix{{\bf1}&{\bf 0}\cr{\bf0}&-{\bf1}\cr}\right).
\label{matrices}
\eeq

\ss{Spinor field inside a spherical shell: bag boundary conditions}

For bag ---also called mixed--- boundary conditions, we apply $P_+\psi=0$ at
$r=1$, where the projection is given by
\beq
P_+={1\over2}\big({\bf 1}-i\Ga^5\Ga^\mu\,n_\mu\big),
\label{mproj}
\eeq
in terms of the inward normal $n_\mu$.
For the geometry of the ball
$$
P_+={1\over2}\left(\matrix{{\bf1}&i{\bf1}\cr-i{\bf1}&{\bf1}\cr}\right) ,
$$
and so for $\psi^{(+)}_\pm$,
$$
J_{n+D/2}(k)=\mp J_{n+D/2-1}(k) ,
$$
and for $\psi^{(-)}_\pm$,
$$
J_{n+D/2-1}(k)=\mp J_{n+D/2}(k),\quad n=0,1,2,\ldots.
$$
Thus, taking $\nu_n=n+(D-2)/2$, the implicit eigenvalue equation is as in
\cite{eath91-43-3234}
\beq
J_\nu^2(k)-J_{\nu +1}^2(k)=0,
\label{mimp}
\eeq
while the degeneracies are
\beq
d(n)=d_s \left(
\begin{array}{c}
 D+n-2   \\
  D-2
\end{array}
\right) .
\label{mdegen}
\eeq
In two dimensions the degeneracy is just $2$.

In summary, all the relevant functions for this case are
\beq
\nu_n=n+\frac{D-2}2\:,\nn
\label{deg12}
\eeq
\beq
d(n)=d_s\frac{(n+D-2)!}{n!(D-2)!}\:,\hs\hs d(n)=2\:,
\mbox{ for }D=2\:,
\eeq
\beq
\zb(s)=d_s\zeb(2s,D/2-1;d)\:,
\hs\hs\zb(s)=2\ze_R(2s)\:,\mbox{ for }D=2\:,\nn
\label{deg12b}
\eeq
\beq
P(\nu,z)=I^2_\nu(z)+I^2_{\nu+1}(z)\:,\nn
\eeq
\beq
F(\nu,z)=\frac{(1-t)e^{2\nu\eta}(1+z^2)^{\frac12}}{\pi\nu z^2}\:,\nn
\eeq
\beq
\ln\left(\frac1{2(1-t)}
\left[\Sigma_1^2+\Sigma_2^2-2t\Sigma_1\Sigma_2\right]\right)
\sim\sum_{n=1}^{\infty}\frac{D_n(t)}{\nu^n}\:,\nn
\eeq
\beq
A_{-1}(s)&=&\frac1{2\sqrt\pi}
\frac{\Gamma\left(s-\frac12\right)}{\Gamma(s+1)}\:
\zb\left(s-1/2\right),\nn\\
A_0(s)&=&-\frac1{2\sqrt\pi}
\frac{\Gamma\left(s+\frac12\right)}{\Gamma(s+1)}\:\zb(s)\nn,
\eeq
with $\ze_R$  the Riemann $\zeta$-function.

The contribution of $\nu=0$, which we have in two dimensions reads here
\beq
Z_0(s)&=&\frac{\sin(\pi s)d[0]}{\pi}\left\{
     \int_0^1dz \,\,z^{-2s}
\frac{\partial}{\partial z}
\ln(I_0^2(z)+I_1^2(z))\right.\nn\\
   &&\hs +\int_1^\infty dz \,\,z^{-2s}
\left[\frac{\partial}{\partial z}
\ln(I_0^2(z)+I_1^2(z))
-2+\frac1{z}-\frac1{4z^2}\right]
   \nn\\
 & &\left.\hs\hs
-\frac1{2s}+\frac1{(s-1/2)}+\frac1{8(s+1/2)}\right\}.
\nn
\eeq

\ss{Spinor field outside a spherical shell: bag boundary conditions}

As in the scalar cases, we must simply replace Bessel with Hankel functions.
Eqs.~(\ref{deg12}) and (\ref{deg12b}) provide some quantities needed in
the computation, while for the rest, we get
\beq
P(\nu,z)=K^2_\nu(z)+K^2_{\nu+1}(z)\:,\nn
\eeq
\beq
F(\nu,z)=\frac{4\nu(1+t)e^{-2\nu\eta}(1+z^2)^{\frac12}}{\pi z^2}\:,\nn
\eeq
\beq
\ln\left(\frac1{2(1+t)}\left[\Sigma_3^2+\Sigma_4^2+2t\Sigma_3
\Sigma_4\right]\right)
\sim\sum_{n=1}^{\infty}\frac{D_n(t)}{\nu^n}\:,\nn
\eeq
\beq
A_{-1}(s)&=&-\frac1{2\sqrt\pi}
\frac{\Gamma\left(s-\frac12\right)}{\Gamma(s+1)}
\:\zb\left(s-1/2\right),\nn\\
A_0(s)&=&\frac1{2\sqrt\pi}
\frac{\Gamma\left(s+\frac12\right)}{\Gamma(s+1)}\:\zb(s)\nn.
\eeq
In the same way, for $\nu=0$ we obtain
\beq
Z_0(s)&=&\frac{\sin(\pi s)d[0]}{\pi}\left\{
     \int_0^1dz \,\,z^{-2s}
\frac{\partial}{\partial z}
\ln(K_0^2(z)+K_1^2(z))\right.\nn\\
   &&\hs +\int_1^\infty dz \,\,z^{-2s}
\left[\frac{\partial}{\partial z}
\ln(K_0^2(z)+K_1^2(z))
+2+\frac1{z}+\frac1{4z^2}\right]
   \nn\\
 & &\left.\hs\hs
-\frac1{2s}-\frac1{(s-1/2)}-\frac1{8(s+1/2)}\right\}.
\nn\eeq

The numerical results for spin $1/2$ with bag boundary conditions are
given in Table~3.
\BagBC
The $D=3$ result is the one found already by Milton \cite{milt83-150-432}
(albeit with far less precision).

\ss{Spinor field with global spectral boundary conditions}
We shall  now obtain the results for spectral boundary conditions
\cite{dowk96-13-2911,dowk99-}.
Such boundary conditions are imposed by setting equal to zero, at $r=1$,
the negative (resp. positive) $Z$-modes
of the positive (resp. negative) chirality parts of $\psi$,
the rest of the modes remaining free.

Roughly speaking, spectral conditions amount to
requiring that zero-modes of (\ref{Dirac}) should be
square-integrable on the elongated manifold obtained from the
ball by extending the narrow collar (of the approximate
product metric $dr^2+d\Om^2$)
just inside the surface, to values of $r$ ranging from $1$ to $\infty$. This
will be so if the modes of $A=\Ga^r\Ga^a\nabla_a$ with {\it negative}
eigenvalues are suppressed at the boundary
(e.g.
\cite{atiy75-77-43}
-\cite{gilk75-15-334}).

From (\ref{matrices}) and (\ref{spheig}), the boundary operator
is $A_0=\Ga^r\Ga^a\nabla_a\big|_{r=1}$
and has for eigenstates
\beq
A_0    \left( \begin{array}{c}
Z^{(n)}_+  \\
Z^{(n)}_-\end{array} \right)
=\la_n
\left( \begin{array}{c}
Z^{(n)}_+   \\
Z^{(n)}_- \end{array} \right)
,\quad
A_0  \left( \begin{array}{c}
Z^{(n)}_-\\
Z^{(n)}_+  \end{array} \right)
=-\la_n   \left( \begin{array}{c}
Z^{(n)}_-   \\
Z^{(n)}_+   \end{array} \right) .
\label{eq2.62}
\eeq
Thus, from (\ref{diracmodes}) we see that the negative modes of $A_0$ are
associated with the radial factor $J_{n+D/2-1} (kr)$. Taking $\nu$ as
before, $\nu=n+(D-2)/2$, the implicit eigenvalue equation reads
\beq
J_\nu(k)=0.\nn
\eeq
The degeneracy for each eigenvalue is
\beq
d(n)=2d_s
\left(
\begin{array}{c}
n+D-2 \\
D-2
\end{array}
\right)\:,
\hs\hs
d(n)=4\:,\mbox{ for }D=2\:.
\label{spi2}
\eeq
The relevant boundary zeta function reads now
\beq
\zb(s)=2d_s\zeb (2s,D/2-1;d)\:,\hs\hs
\zb(s)=4\ze_R(2s)\:,\mbox{ for }D=2\:.
\label{spi4}
\eeq
As we see, apart from the degeneracy of the eigenvalues and the relation between
$\zb$ and the Barnes $\zeta$-function,  the rest of the argumentation
is identical to that for the
scalar case with Dirichlet boundary conditions.
Thus, Eqs.~(\ref{f0D})-(\ref{Z0sc}) remain valid once
the above definitions are used.

For the exterior space we have to employ Eqs.~(\ref{spi2}) and (\ref{spi4})
in Eqs.~(\ref{p0De1})-(\ref{p0De2}), but it has to be noted that here,
in contrast with the interior case, $\nu_l=l+D/2$,
as a result of the normal vector changing its sign.
This means that there is no $\nu_l=0$ contribution.

The numerical results for this case are listed in Table~4, for
$D=2,...,9$.
\GSBC

\s{Electromagnetic field in a perfectly conducting spherical shell}
The Casimir energy of the electromagnetic field is, essentially, the sum of
a Dirichlet and of a Robin scalar field (with a specific value for $\beta$,
see Eq. (\ref{cas3.1a})), the only difference being that the angular momentum
$l=0$ is to be omitted.
An exception is $D=2$, where the vector Casimir effect consists of only the
transverse magnetic mode contributions.
Being precise, in the interior of the shell one has for the
transverse electric (TE) ---respectively for the transverse magnetic (TM)
modes--- the
following boundary conditions \cite{boye68-174-1764,slat47b},
\beq
\begin{array}{cc}
r^{1-D/2}J_{\nu_l} (\om_{l,n} r) |_{r=R} =0, & \mbox{for TE-modes,} \\
\left[(D/2-1)J_{\nu_l} (\om_{l,n} r)+\om_{l,n}J'_{\nu_l} (\om_{l,n} r)\right]
|_{r=R} =0, & \mbox{for TM-modes} .\label{cas9.1}
\end{array}
\eeq
The condition for the TM-modes is of Robin type with $\beta=2-D$.
Since the $l=0$ mode has to be omitted, the minimum eigenvalue in this case
is $\mu_1=D/2$ and therefore we can apply the method for any $\be=2-D>1-D$.
Thus, in order to get the Casimir energy of the electromagnetic field, we
must simply  repeat the computation of Sect.  \ref{S:scalar}
for Dirichlet and Robin boundary conditions with $\be=2-D$
and add them up.
We have to exclude everywhere the $l=0$ mode and this means
that also the base $\zeta$-function is slightly modified,
in the way
\beq
\zb(s)=\zeb\at2s,\frac{d+1}2;d\ct+\zeb\at2s,\frac{d-1}2;d\ct
-\at\frac{d-1}2\ct^{-2s}\:.
\eeq
The results for the electromagnetic field are summarised in Table~5.
\EF
$D=2$ is the Neumann result, $D=3$ is the well known figure first
obtained by Boyer \cite{boye68-174-1764} and later recalculated in Refs.
\cite{milt78-115-388,bali78-112-165}.

\section{Discussion and conclusions}
In Ref. \cite{bord96-37-895}, two of the present authors
 developed a new, seminal approach for finding
representations of the zeta function associated with the Laplace
operator on the $D$-dimensional ball. At that stage, dimension by
dimension was considered, but soon a refined and generalised technique was
provided in subsequent works
\cite{bord96-182-371,dowk99-}. Making use of the Barnes zeta function
\cite{barn03-19-374}, dimension can easily be dealt with as a parameter
and several different fields can also be treated on the same footing.
The representations derived
are valid for all values of the complex parameter $s$ and it is up to
the practitioner's needs or wishes at which values of $s$ the zeta
function is to be evaluated. In previous work
our concern was of a more mathematical nature and we considered function
values and residues appropriate to find  heat-kernel coefficients
\cite{bord96-37-895,bord96-182-371,dowk99-}, as well as the derivative at
$s=0$ \cite{bord96-179-215}. Since then, a number of proclaimed ``new"
methods have been developed in the literature.

 Our aim in the present article has been to show explicitly that Casimir
energies for the big family of the more usual configurations
can be obtained in fact from {\it general} formulas, also
in quite non-trivial situations,
 where the boundaries are not  flat plates, the fields are
spinorial (rather than scalar), and also when the
 boundary conditions are very general and rather involved.
We have gone much beyond previous work in that we are not here restricted
 any more  to a very specific field in a specific
dimension with a specific boundary condition, but give general formulas for
 basically any
possible situation that can arise in practice, involving
spherically symmetric boundaries.

Some comments on the precision and accuracy of the numerical procedure
employed are in order. It is clear from the analysis developed in the previous
sections, that a
numerical evaluation of the asymptotic terms $A_i (s)$ to {\it any}
 desired accuracy is immediate, using the formulas given there.
These contributions are always represented by known
special functions and using available programs, such as Mathematica,
the accuracy with
which these are calculated is readily obtained. Imposing accuracies of, e.g.,
$10^{-20}$ or more, we get results in negligible cpu time.

The only problem (if at all) with the numerical analysis is the computation
of $Z(s)$, Eq.~(\ref{PF}). It is twofold. On one hand,
 the integration, up to {\it infinity}, of  the combination
of Bessel functions is not strictly
possible, using the exact form of the Bessel functions. On the other hand, the
angular summation, up to {\it infinity}, cannot
be performed exactly. For large angular momenta,  the Bessel functions
take a rather complicated form, what renders exact summation not possible.
For that reason, the following procedure has been applied throughout
(the error bounds given below are for Dirichlet
boundary conditions, but very similar relations hold for the other conditions
considered).

We have dealt with the infinite integration as follows. The main
contributions always come from small values of $z$, and thus we split
the integral into
$$
\int_0^B dz + \int_B^\infty dz.
$$
Whereas in the first integral the Bessel functions themselves are used
for the integration, in the second integral their asymptotic expansion
for large arguments is employed. The value of $B$ is computed with the help
of an adaptative procedure, such that the integrand and its asymptotic
expansion differ, at $B$,
by less than, say $10^{-12}$. Typically $B=10$ is already sufficient. Given
that the asymptotics of the Bessel functions is a simple polynomial in powers
of $(1/z)$, the integration up to infinity is very easily done.

Let us now assume that the contribution of the first $L$ angular momenta has
been calculated as described. In order to get a numerical approximation
for the angular momentum sum, from $L+1$ to infinity, we proceed
as follows. The idea is that, for sufficiently large values of $L$,
the integrand can be replaced by its uniform asymptotic expression.
For Dirichlet boundary conditions this amounts to going through the following
steps:
\begin{eqnarray}
Z^{int}_{L+1}& \equiv &- \frac 1 \pi \sum_{l=L+1}^\infty d(l) \nu
            \int_0^\infty dz \,\,
      \left[ \ln I_{\nu} (\nu z) -\ln \frac{e^{\nu \eta} }
          {\sqrt{2\pi\nu} (1+z^2)^{1/4}} -\sum_{n=1}^N \frac{D_n (t)}{\nu^n}
            \right] \nn\\
   &\sim & -\frac 1 \pi \left[ \left( \int_0^\infty dz \,\, D_{N+1} (t)\right)
                       \sum_{l=L+1}^\infty d(l) \nu ^{-N} \right.\nn\\
   & &\left.+
               \left( \int_0^\infty dz \,\, D_{N+2} (t)\right)
                     \sum_{l=L+1}^\infty d(l) \nu ^{-N-1} +... \right]\nn \\
  &=& -\frac 1 \pi \left[ \left( \int_0^\infty dz \,\, D_{N+1} (t)\right)
        \left( \zeta_{{\cal N}} (N/2) -\sum_{l=0}^L d(l) \nu^{-N} \right)
              \right.\nn\\
  & & \left.
                 +\left( \int_0^\infty dz \,\, D_{N+2} (t)\right)
        \left( \zeta_{{\cal N}} ((N+1)/2) -\sum_{l=0}^L d(l) \nu^{-N-1}\right)
       +...\right] . \label{numpro1}
\eeq
Again, the integrals over the uniform asymptotics are simple and can be
performed analytically. In this way, a closed expression for the approximation
is found, the value of $L$ being again determined by an adaptative procedure.
By definition, the difference $Z^{int}_L - Z^{int}_{L+1}$ equals
the contribution coming
from $l=L$. The value of $L$ is determined such that the difference
$Z^{int}_L - Z^{int}_{L+1}$, obtained from (\ref{numpro1}), agrees up to
say $10^{-10}$ with the contribution from $l=L$ calculated
previously. Depending on the dimension, the values of $L$ range from
6 (for D=9) to 49 (for D=3).

In summary, as explained, this procedure takes fully into account
the integrals of infinite range as well as the
summation up to infinity. The error bounds can thus be imposed at will
in the single steps and this
guarantees that the results given are always numerically precise, up
to {\it any} pre-established digit. To our knowledge, this does not apply
to any other method.

In the cases when partial results were known, we have compared our numbers
 with those while improving always, by several digits, such
known values and deriving, for the first time, a lot of new ones, for
different fields (e.g. results for the exterior space in the case of the
electromagnetic field)
and different boundary conditions (e.g. for spectral boundary conditions, and
 for bag boundary conditions in any dimension).
For the scalar field with Dirichlet boundary conditions we have re-obtained,
in particular, the
known result that for $D$ even the energy is divergent
\cite{bend94-50-6547}. Here it still remains unclear if there may be a
natural way to get, unambiguously, a finite answer with physical sense
for this case. In  odd dimension,
$D=2n-1$, the sign of the Casimir energy seems to be determined by the sign
of $(-1)^n$. For even dimension, $D=2n$, one also finds the alternating
structure $(-1)^{n+1}$ for the {\it finite} part of the Casimir energy; however,
its interpretation is unclear due to the presence of the pole.
Similar comments hold for the interior and exterior contributions
separately, with the same problems of interpretation. For
Neumann boundary conditions, in all the dimensions calculated,
the Casimir energy is negative.
Similarly, one can describe the results summarized in Tables 3--5.
In all cases we have been able to obtain {\it general}, {\it highly accurate}
 expressions which,
by fixing some parameters, provide us with the desired specific
example and yield a numerical answer of arbitrary precision
(just by adding the convenient number of terms of the corresponding series).

Disappointing as the mentioned ---quite well known--- ambiguities may be
(specialists in the field are quite used to them by now), even more so is the
 fact that no general pattern seems to arise from our general formulas
which might hint towards
 the physical understanding of the final {\it sign} of the energy.
Here we have been able to demonstrate, without reasonable doubt,
 the existence of
the two classes of Casimir force: attractive and repulsive, but are unable to
give the rule according to which one or the other will show up at a
particular instance. Further insight will be needed to clarify this point.

\vspace{3mm}

\noindent{\large \bf Acknowledgments}

Thanks are given to the referees for very constructive remarks.
 This investigation has been  supported by
 DGICYT (Spain), project
PB96-0925, by DFG under contract number BO1112/4-2, by the
German-Spanish Program Acciones Integradas, project HA1997-0053
 and by the Program CICYT(Spain)-INFN(Italy).

\end{document}